# Ghost-wave momentum bandgaps in anisotropic Floquet lattices

Junhua Dong, Huan He, and Huanan Li*

*MOE Key Laboratory of Weak-Light Nonlinear Photonics, School of Physics, Nankai University, Tianjin 300071, China*

*Momentum bandgaps, characterized by complex frequencies and non-resonant amplification effects, provide a powerful route for wave manipulation beyond conventional band theory. Here, we introduce a distinct mechanism for momentum-gap engineering in higher-dimensional Schrödinger-type Floquet lattices by exploiting the intrinsically complex wave vectors of ghost waves, with complex-frequency excitation providing an additional degree of freedom for continuously tailoring the ghost-wave branch and the associated Floquet spectrum. Furthermore, we show that the higher-dimensional Floquet band structure supports momentum bandgaps extending across the entire Brillouin zone along the propagation direction and enables amplification over a broad frequency range under arbitrarily weak modulation. When the lattice is truncated along the ghost-wave decay direction, the resulting Floquet waveguide exhibits broadband reflectionless pulse amplification. Our results establish a higher-dimensional framework for ghost-wave momentum-gap physics and reveal new opportunities for non-Bloch wave engineering in time-varying photonic systems.*

Photonic time crystals (PTCs), formed by periodically modulating a uniform medium in time, constitute the temporal counterpart of photonic crystals and have opened new opportunities for controlling wave propagation [1]-[7]. As a manifestation of space-time duality, PTCs support temporal scattering in which momentum, rather than frequency, is conserved [8]. The resulting time-refracted and time-reflected waves interfere along the propagation direction, giving rise to a band structure in momentum space featuring momentum bands and momentum bandgaps. Drawing energy from the temporal modulation, modes within a momentum bandgap can grow exponentially in time, enabling a variety of exotic phenomena, such as nonresonant tunable lasing [9]-[13] and topological temporal edge states [14]-[17]. However, opening an appreciable momentum bandgap in PTCs generally requires strong and ultrafast modulation [4],[18], posing a formidable challenge for optical implementations and restricting most experimental demonstrations to the microwave and acoustic domains [15],[19]-[23]. To overcome this limitation, considerable effort has been devoted to enlarging momentum bandgaps under weak-modulation conditions, notably by exploiting strong resonant effects [24]-[27], nonuniform waves [28], and material nonlocality [29].

* hli01@nankai.edu.cn

Recently, Floquet lattices realized across diverse wave platforms have emerged as a new setting for momentum-gap physics and temporal topology [30]-[32]. A fundamental distinction from PTCs is that their dynamics are governed by first-order Schrödinger-like evolution equations (in continuous or discrete time), rather than by the second-order electromagnetic wave equation [14],[33]. Due to this distinction, and because Hermitian Schrödinger-type Floquet systems do not support momentum bandgaps, existing realizations typically rely on non-Hermitian gain-loss engineering [30]. Accordingly, the physical origin of the momentum gap and its associated amplification can differ from that in PTCs, where both arise directly from temporal modulation. An extreme example is the recent observation of topological temporal boundary states in a non-Hermitian spatial crystal subjected to a temporal gain-loss flip, where the gap modes originate entirely from the parity-time-broken phase of the underlying static crystal rather than from the temporal modulation itself [34]. Despite these recent advances, momentum bandgaps achievable in Floquet lattices remain limited in both momentum and frequency, much like in conventional PTCs. Whether large momentum bandgaps can be realized solely through weak temporal modulation, without material gain and in a manner intrinsic to Floquet dynamics, remains elusive.

In this work, we propose a distinct route toward momentum-gap physics based on higher-dimensional Floquet lattices, going beyond the conventional framework of space-time duality. Specifically, we extend ghost-wave momentum-gap physics to two-dimensional (2D) anisotropic Floquet lattices that support ghost waves with complex wave vectors, exhibiting simultaneous oscillatory and evanescent behavior. By exploiting these complex wave vectors, momentum gaps can be engineered without introducing material gain, while additional tunability is achieved through complex-frequency excitation, revealing band structures unattainable with real wave vectors. We further demonstrate that, in a waveguide geometry formed by truncating the Floquet lattice along the ghost-wave decay direction, the ghost-wave momentum bandgap extends across the entire Brillouin zone and a broad frequency range even under arbitrarily weak modulation, leading to broadband pulse amplification without back-reflection. Our approach uncovers an exotic mechanism for momentum-gap engineering rooted in the asymmetry between spatial and temporal dimensions in higher-dimensional Floquet lattices.

***Anisotropic Floquet lattices*** — In photonic platforms, tight-binding lattice systems can be realized using coupled-resonator arrays, synthetic frequency dimensions, or combinations of both [35]. In such lattices, each site corresponds to a discrete optical mode, and its amplitude represents the complex field amplitude of that mode. The resulting dynamics are described by a Schrödinger-like equation within the framework of temporal coupled-mode theory [36]-[38]. Here, we consider a 2D non-Hermitian Su-Schrieffer-Heeger (SSH) lattice composed of SSH columns. The columns are coupled uniformly along the $x$ direction with nearest-neighbor coupling $t_0 > 0$. For an isolated column, the intra-column dynamics along the $y$ direction are described by the non-Hermitian SSH Hamiltonian

$$H_y = \sum_n [t_1 |n, A\rangle\langle n, B| + t_{-1} |n, B\rangle\langle n, A| + t_2(|n+1, A\rangle\langle n, B| + |n, B\rangle\langle n+1, A|)],$$

where $t_1 = t_0 e^g$ and $t_{-1} = t_0 e^{-g}$ are nonreciprocal intracell couplings, while $t_2 = t_0$ is the reciprocal intercell coupling. Without loss of generality, the onsite terms are taken to be zero by an appropriate choice of the reference frequency. By periodically switching the anisotropy parameter $g$ between a nonzero value $g^{(1)}$ and $g^{(2)} = 0$, we obtain an anisotropic Floquet lattice (AFL) with a time-periodic column Hamiltonian $H_y(t) = H_y(t+T)$, where $T$ is the modulation period [see Fig. **1**(a)].

In the site basis, $H_y(t)$ acts on the column field-amplitude vector $\Psi_m(t) = \left[\cdots, \quad \psi_{-1,m}^T(t), \quad \psi_{0,m}^T(t), \quad \psi_{1,m}^T(t), \quad \cdots\right]^T$, where the superscript $T$ denotes the transpose, and $\psi_{n,m}(t) = [\psi_{n,m}^A(t), \quad \psi_{n,m}^B(t)]^T$ collects the field amplitudes on the $s = A, B$ sublattices of the $n$th unit cell in the $m$th column. The dynamics of the full 2D Floquet lattice are governed by

$$i\frac{d\Psi_m}{dt} = t_0\Psi_{m-1} + H_y(t)\Psi_m + t_0\Psi_{m+1}. \tag{1}$$

Since the coupling strengths determine the characteristic energy scale of the lattice and are typically orders of magnitude smaller than the underlying optical resonance frequency, the relevant modulation frequency $\omega_0 = 2\pi/T$ and switching rate are chosen relative to the coupling scale rather than the optical carrier frequency, allowing temporal modulation at rates far below the optical oscillation frequency [39]-[40]. Experimentally, the two lattice dimensions can be implemented by combining a physical resonator lattice with a synthetic frequency dimension, where the indices $m$ and $ns$ denote the real-space and synthetic-space coordinates, respectively. Electro-optic modulation provides dynamic control over synthetic-frequency coupling, enabling the realization of non-Hermitian anisotropic coupling through phase and amplitude modulation [41]. Such modulation schemes also offer a promising route to temporally engineer the anisotropic synthetic dimension and implement the proposed Floquet switching protocol.

Owing to the translational invariance along the $x$ direction, the Floquet lattice admits coordinate-separable solutions of the form $\psi_{n,m}(t) = Y_n(t)\exp(ik_x m - i\omega_x t)$, with $\omega_x = 2t_0 \cos k_x$. Substituting this ansatz into Eq. (1) reduces the problem to

$$i\frac{dY_n}{dt} = \begin{bmatrix} 0 & t_0 e^{g(t)} \\ t_0 e^{-g(t)} & 0 \end{bmatrix} Y_n + \begin{bmatrix} 0 & 0 \\ t_0 & 0 \end{bmatrix} Y_{n+1} + \begin{bmatrix} 0 & t_0 \\ 0 & 0 \end{bmatrix} Y_{n-1}, \quad (2)$$

which governs the dynamics of the reduced unit-cell amplitude $Y_n(t)$. Notably, Eq. (2) for $Y_n(t)$ is independent of the wavevector component $k_x$, and hereafter, we assume a real $k_x \in (-\pi, \pi)$, corresponding to propagating modes along the $x$ direction.

***Ghost waves in anisotropic lattices: from real- to complex-frequency excitation***—Ghost waves were originally introduced in transparent anisotropic dielectric media as modes with complex wave vectors that combine the properties of propagating and evanescent fields [42]-[44]. More recently, this concept was extended to anisotropic PTCs for momentum-bandgap engineering [28]. Here, we first consider the lattice counterpart of ghost waves in a static anisotropic lattice with a fixed anisotropy parameter $g(t) = g$. For the static lattice, we seek Bloch solutions of the form $Y_n(t) = v\exp(ik_y n - i\omega_y t)$ along the $y$ direction, under which Eq. (2) reduces to the eigenvalue problem

$$H(k_y)v = \omega_y v, \quad (3)$$

with the Bloch Hamiltonian $H(k_y) = \begin{bmatrix} 0 & h_+ \\ h_- & 0 \end{bmatrix}$, where $h_\pm = (e^{\pm g} + e^{\mp i k_y}) t_0$. Solving the characteristic equation $\det[H(k_y) - \omega_y I_2] = 0$ (where $I_2$ is the $2 \times 2$ identity matrix) yields the dispersion relation

$$e^{ik_y} = \frac{-2t_0^2 + \omega_y^2 \pm \sqrt{\Delta}}{2t_0^2} e^{-g}, \tag{4}$$

with discriminant $\Delta = \omega_y^4 - 4t_0^2\omega_y^2$. As the angular frequency $\omega_y$ varies, two distinct wave types arise: evanescent waves for $\Delta > 0$, where $k_y$ is purely imaginary, and ghost waves for $\Delta < 0$, where $k_y = k_y' + ik_y''$ has simultaneously nonzero real and imaginary parts, corresponding to oscillatory and exponentially decaying behavior. The latter occur in the frequency range $-2t_0 < \omega_y < 2t_0$ and constitute the main focus of the present work.

Beyond varying the frequency $\omega_y$, an additional degree of freedom for tuning $k_y$ is provided by complex-frequency excitation. As the system approaches a quasi-stationary state oscillating at the same complex frequency as the excitation, effective loss or gain emerges depending on whether the signal amplitude grows or decays exponentially in time [45]-[49]. This, in turn, enables continuous modulation of $k_y$. Indeed, by extending the excitation frequency from the real axis $\omega_y \in \mathbb{R}$ to complex frequencies $\omega_y = \omega_y' + i\omega_y''$ with $\omega_y' > 0$, the dispersion relation in Eq. (4) can be rewritten as

$$\omega_y' = 2t_0 \cos\frac{k_y'}{2} \cosh\frac{g - k_y''}{2}, \qquad \omega_y'' = 2t_0 \sin\frac{k_y'}{2} \sinh\frac{g - k_y''}{2}. \tag{5}$$

For each $\omega_y$, Eq. (5) admits two physically distinct complex wavevector solutions $k_{y\pm} = k_{y\pm}' + ik_{y\pm}''$, related through $k_{y-}' = -k_{y+}'$ and $k_{y-}'' = 2g - k_{y+}''$. For ghost waves with $\omega_y'' = 0$, the attenuation constant satisfies $k_y'' = g$.

The physical connection between the attenuation constant $k_y''$ and the temporal growth rate $\omega_y''$ of the excitation can be understood by examining the energy flux $J_{n\leftarrow n-1}$ between adjacent unit cells along the $y$ direction. For the eigenmode $Y_n(t) = \begin{bmatrix} \xi \\ 1 \end{bmatrix} \exp(ik_y n - i\omega_y t)$, obtained from Eq. (3), with $\xi = \omega_y / h_-$, the energy flux in the quasi-stationary state takes the form [50]

$$J_{n\leftarrow n-1} = -2t_0 \sin\frac{k_y'}{2} e^{\frac{g+3k_y''}{2}} e^{2\omega_y'' t - 2k_y'' n}.$$

Unlike the ghost waves originally proposed in anisotropic dielectrics, the energy flux $J_{n\leftarrow n-1}$ remains nonzero even for $\omega_y'' = 0$, owing to the non-Hermitian coupling in our lattice system. As an illustrative example, for $k_y' < 0$, the ghost wave carries energy along the positive $y$ direction, i.e., $J_{n\leftarrow n-1} > 0$, with attenuation factor $\exp(-2gn)$. Introducing virtual loss ($\omega_y'' > 0$) or gain ($\omega_y'' < 0$) increases or decreases, respectively, the attenuation factor $\exp(-2k_y'' n)$ of $J_{n\leftarrow n-1}$, as follows directly from Eq. (5).

We focus on the branch of the complex wave vector $k_y = k_y' + ik_y''$ satisfying $k_y'' \leq g$. For each fixed $k_y''$, the excitation frequency $\omega_y = \omega_y' + i\omega_y''$ traces an elliptical trajectory in the complex-frequency plane as $k_y'$ varies over $(-\pi, \pi)$. According to Eq. (5), this trajectory obeys $(\omega_y')^2/a_1^2 + (\omega_y'')^2/a_2^2 = 1$, with the semi-axis lengths $a_1 = 2t_0 \cosh[(g - k_y'')/2]$ and $a_2 = 2t_0 \sinh[(g - k_y'')/2]$. During the initial modulation stage, the AFL reduces to the static anisotropic lattice with anisotropy parameter $g = g^{(1)}$ [see Fig. **1**(a)]. In Fig. **1**(b), we plot these elliptical trajectories for $g^{(1)} = 0.1$, $t_0 = 0.3\omega_0$, and for the representative values $k_y'' = g$, $0.75g$, $0.5g$, and $0.25g$. The special case $k_y'' = g$ corresponds to real-frequency excitation ($\omega_y'' = 0$), whereas decreasing $k_y''$ requires progressively larger values of $|\omega_y''|$ for a fixed $\omega_y'$. Fig.

**1**(c) further shows $\mathrm{Im}\,\omega_y^{(1)}$ as a function of $k_y'$, illustrating that complex-frequency excitation provides an additional tuning knob for engineering the ghost-wave branch.

***Band structure of AFLs for ghost waves***—In infinite AFLs alternating between anisotropic ($q = 1$) and isotropic ($q = 2$) configurations, the preserved translational symmetry ensures that both $k_x$ and $k_y$ remain conserved throughout the Floquet modulation. For each fixed $k_y$, the static lattice with anisotropy parameter $g$ supports two eigenmodes along the $y$ direction, $Y_{n\pm}(t) = v_\pm \exp\left(ik_y n - i\omega_{y\pm} t\right)$, obtained from Eq. (3), with opposite eigenfrequencies $\omega_{y\pm} = \pm\omega_y$ and corresponding eigenvectors $v_\pm = \begin{bmatrix} \pm\xi \\ 1 \end{bmatrix}$, where $\xi = \omega_y / h_-$ and $\omega_y$ is given by Eq. (5). Using these eigenmodes for $g = g^{(q)}$ in the $q = 1,2$ modulation stages, the reduced unit-cell amplitude $Y_n(t)$ satisfying Eq. (2) can be expressed as

$$Y_n(t) = \left(A_+(t) v_+^{(q)} + A_-(t) v_-^{(q)}\right) e^{ik_y n}, \tag{6}$$

where $q$ denotes the modulation stage containing time $t$, and the mode amplitudes evolve as $A_\pm(t) \propto e^{-i\omega_{y\pm}^{(q)} t}$ within each stage.

The evolution of the amplitude $Y_n(t)$ in Eq. (6) throughout the AFL can be determined using the temporal transfer-matrix approach. Specifically, defining the state vector $A(t) = \begin{bmatrix} A_+(t) \\ A_-(t) \end{bmatrix}$, its free evolution during the interval $T^{(q)}$ in the $q = 1,2$ stages is governed by the diagonal propagation matrix $F^{(q)} = \mathrm{diag}\left\{e^{-i\omega_{y+}^{(q)} T^{(q)}}, e^{-i\omega_{y-}^{(q)} T^{(q)}}\right\}$. At a temporal interface when the lattice transitions from stage $q = 1$ to $q = 2$, continuity of $Y_n$ across the interface induces a discontinuity in the state vector $A$, governed by matching matrix

$$J^{(2,1)} = \frac{1}{2\xi^{(2)}} \begin{bmatrix} \xi^{(2)} + \xi^{(1)} & \xi^{(2)} - \xi^{(1)} \\ \xi^{(2)} - \xi^{(1)} & \xi^{(2)} + \xi^{(1)} \end{bmatrix}.$$

Conversely, at the reverse temporal interface when the lattice switches from $q = 2$ back to $q = 1$, the matching matrix becomes $J^{(1,2)} = \left[J^{(2,1)}\right]^{-1}$. The combination of the propagation and matching matrices dictates the evolution of the state vector $A(t)$ in the AFL. The resulting transfer matrix for a temporal unit cell of the AFL, indicated by the red dashed box in Fig. **1**(a) and having duration $T = T^{(1)} + T^{(2)}$, is given by $M(k_y) = J^{(1,2)}F^{(2)}J^{(2,1)}F^{(1)}$.

The eigenvalues $\lambda_\pm$ of the temporal-unit-cell transfer matrix $M(k_y)$, i.e., $M(k_y)|\pm\rangle = \lambda_\pm|\pm\rangle$, are related to the Floquet frequency $\Omega_{y\pm}$ through $\lambda_\pm = e^{-i\Omega_{y\pm}T}$. By solving the secular equation $\det\{M(k_y) - e^{-i\Omega_{y\pm}T}I_2\} = 0$, we obtain the band structure, namely $\Omega_{y\pm}$ versus $k_y$, associated with the Floquet dynamics of the reduced amplitude $Y_n(t)$ along the $y$ direction. Noting that $\det M(k_y) = 1$ and choosing $\mathrm{Re}\,\Omega_{y\pm} \in (-\omega_0/2, \omega_0/2)$ without loss of generality, the Floquet frequencies $\Omega_{y\pm}$ are opposite in sign, i.e., $\Omega_{y+} = -\Omega_{y-}$. Furthermore, since $M(-k_y^*) = \begin{bmatrix} 0 & 1 \\ 1 & 0 \end{bmatrix}^{-1} M(k_y)^* \begin{bmatrix} 0 & 1 \\ 1 & 0 \end{bmatrix}$, the Floquet frequencies at $-k_y^*$ and $k_y$ satisfy $\Omega_{y\pm}(-k_y^*) = -\Omega_{y\pm}(k_y)^*$. Bearing this symmetry in mind, we restrict our discussion to the band structure of the AFL for $k_y' \in (0, \pi)$.

Associated with the specific values of $k_y''$ considered in Fig. **1**(b, c), Fig. **2**(a) plots the complex Floquet frequencies $\Omega_{y\pm}$ as functions of $\mathrm{Re}\,\omega_y^{(1)}$, i.e., the real part of the excitation frequency during the $q = 1$ stage. Interestingly, $\mathrm{Re}\,\Omega_{y\pm}$ is nearly independent of $k_y''$, with all curves overlapping in the upper panel. In contrast, $\mathrm{Im}\,\Omega_{y\pm}$, which determines the momentum-bandgap strength, depends strongly on $k_y''$ (lower panel). For the conventional ghost-wave branch ($k_y'' = g$), corresponding to real-frequency excitation ($\omega_y'' = 0$), the momentum bandgap remains open throughout the entire ghost-wave frequency range (blue curve). Complex-frequency excitation generally preserves the bandgap while continuously modifying its strength $\left|\mathrm{Im}\left(\Omega_{y\pm}\right)\right|$, with the spectra for $k_y'' = 0.75g$ and $0.25g$ coinciding and thus appearing as a single green curve. A notable exception occurs at $k_y'' = 0.5g$ (dark-red curve), where the bandgap is open only over a limited frequency interval and reaches its maximum strength near $\mathrm{Re}\,\omega_y^{(1)} = 0.5\omega_0$. Figure **2**(b) further quantifies this behavior by showing the bandgap strength $\left|\mathrm{Im}\left(\Omega_{y\pm}\right)\right|$ as a function of $k_y''$ for $\mathrm{Re}\,\omega_y^{(1)} = 0.45\omega_0$, $0.5\omega_0$ and $0.55\omega_0$ , as indicated by the vertical dashed lines in Fig. **2**(a). The curves are symmetric about $k_y'' = 0.5g$, where the bandgap strength attains its minimum. At $\mathrm{Re}\,\omega_y^{(1)} = 0.5\omega_0$, this minimum remains nonzero, indicating that the momentum bandgap stays open, whereas it vanishes for $\mathrm{Re}\,\omega_y^{(1)} = 0.45\omega_0$ and $0.55\omega_0$, corresponding to bandgap closure.

To understand the Floquet band structure, in particular the behavior of the special branch $k_y'' = g/2$, we introduce the Floquet Hamiltonian $H_F$ of the infinite AFL [51]-[52]. Substituting the ansatz $Y_n(t) = v(t)\exp\left(ik_y n\right)$ into Eq. (2) yields the time-dependent Schrödinger equation $i\,dv(t)/dt = H\left(k_y, t\right)v(t)$, with effective Hamiltonian $H\left(k_y, t\right) = \begin{bmatrix} 0 & h_+(t) \\ h_-(t) & 0 \end{bmatrix}$, where $h_\pm(t) = \left(e^{\pm g(t)} + e^{\mp ik_y}\right)t_0$ . Expanding the time-periodic Hamiltonian and Floquet state as $H\left(k_y, t\right) = \sum_p \widetilde{H}\left(k_y, p\right)\exp(-ip\omega_0 t)$ and $v(t) = \sum_p \tilde{v}_p \exp\left[-i\left(\Omega_y + p\omega_0\right)t\right]$ yields, upon matching harmonics, the time-independent eigenvalue problem

$$\sum_{l=-\infty}^{+\infty}\left[\widetilde{H}\left(k_y, p-l\right) - p\omega_0\delta_{pl}I_2\right]\tilde{v}_l = \Omega_y\tilde{v}_p, \tag{7}$$

or equivalently, $H_F\left(k_y\right)\tilde{v} = \Omega_y\tilde{v}$, where $H_F$ denotes the Floquet Hamiltonian and $\tilde{v}$ is the Floquet eigenvector consisting of Fourier components $\tilde{v}_p$. For a fixed Floquet frequency $\Omega_y$, the secular equation $\det\left(H_F\left(k_{y,(\sigma)}\right) - \Omega_y I\right) = 0$, with $I$ denoting the identity operator in the Floquet space, admits an infinite set of solutions $k_{y,(\sigma)}$, each corresponding to a distinct Floquet mode indexed by $\sigma$. The associated time evolution of each Floquet mode is

$$Y_{n,(\sigma)}(t) = \sum_{p} \hat{v}_{p,(\sigma)} e^{-i(\Omega_y + p\omega_0)t} \, e^{ik_{y,(\sigma)}n}, \tag{8}$$

where $\hat{v}_{p,(\sigma)} = \tilde{v}_{p,(\sigma)} / \sqrt{\sum_p \left|\tilde{v}_{p,(\sigma)}\right|^2}$ is the normalized Fourier component.

In the weak-modulation limit $g^{(1)} \to g^{(2)} = 0$, the eigenvalue problem in Eq. (7) admits an analytical treatment. Neglecting the weak inter-harmonic coupling $\widetilde{H}(k_y, p-l)$ for $p \neq l$, we diagonalize the average Hamiltonian $\widetilde{H}(k_y, 0) = T^{-1}\int_0^T H(k_y, t)dt$ through the similarity transformation $P^{-1}\widetilde{H}(k_y, 0)P = \mathrm{diag}\{\Omega^0_{y+}, \Omega^0_{y-}\}$, where $P = \begin{bmatrix} \xi^{(0)} & -\xi^{(0)} \\ 1 & 1 \end{bmatrix}$, $\xi^{(0)} = \Omega^0_{y+}/h^{(0)}_-$, and $\Omega^0_{y\pm} = \pm\sqrt{h^{(0)}_+}\sqrt{h^{(0)}_-}$. Here, $h^{(0)}_\pm = \left(h^{(1)}_\pm + h^{(2)}_\pm\right)/2$ are the arithmetic averages of the corresponding matrix elements of the Bloch Hamiltonians $H^{(q)}(k_y)$ in Eq. (3) for the two modulation states $q = 1, 2$. The resulting Floquet spectrum forms a ladder with spacing $\omega_0$, given by $\{\Omega^0_{y+} - p\omega_0, \Omega^0_{y-} - p\omega_0\}$. For the branch $k''_y = g/2$, the unperturbed Floquet frequencies are purely real and become $\Omega^0_{y\pm} = \pm t_0\sqrt{\frac{3+\cosh g^{(1)}}{2} + 2\cos k'_y \cosh\left(\frac{g^{(1)}}{2}\right)}$.

Level crossings occur at specific values of $k'_y$ satisfying $\Omega^0_{y+} = \Omega^0_{y-} + \omega_0$. Under weak modulation, the coupling between nearly degenerate Floquet harmonics lifts the degeneracy and modifies the spectrum in the vicinity of the crossing. Using the similarity transformation defined by $P$, we express the Floquet Hamiltonian $H_F$ in the basis that diagonalizes $\widetilde{H}(k_y, 0)$. Retaining only the nearly degenerate subspace, the Floquet spectrum is described by the reduced eigenvalue equation

$$\begin{bmatrix} \Omega_{y-}^0 + \omega_0 & \widetilde{H}'_{21}(k_y, -1) \\ \widetilde{H}'_{12}(k_y, 1) & \Omega_{y+}^0 \end{bmatrix} \begin{bmatrix} (\tilde{v}'_{-1})_2 \\ (\tilde{v}'_0)_1 \end{bmatrix} = \Omega_y \begin{bmatrix} (\tilde{v}'_{-1})_2 \\ (\tilde{v}'_0)_1 \end{bmatrix}, \quad (9)$$

where $\widetilde{H}'(k_y, \pm 1) = P^{-1}\widetilde{H}(k_y, \pm 1)P$, $\tilde{\varphi}'_p = P^{-1}\tilde{\varphi}_p$, and $\widetilde{H}'_{ij}$ and $(\tilde{\varphi}'_p)_j$ denote the corresponding matrix elements and vector components. Using $\Omega_{y+}^0 = -\Omega_{y-}^0$ and $\widetilde{H}'_{21}(k_y, -1) = \widetilde{H}'_{12}(k_y, 1)$, Eq. (9) yields

$$\Omega_{y\pm} = \frac{1}{2}\left[\omega_0 \pm \sqrt{4\left[\widetilde{H}'_{21}(k_y, -1)\right]^2 + \left(\omega_0 - 2\Omega_{y+}^0\right)^2}\right]. \quad (10)$$

For the special branch $k''_y = g/2$, the off-diagonal coupling element in Eq. (9), $\widetilde{H}'_{21}(k_y, -1) = 2it_0^2\left(\cos k'_y + \cosh(g/2)\right)\sinh(g/2)/\pi\Omega_{y+}^0$, is purely imaginary. Consequently, Eq. (10) yields complex Floquet frequencies $\Omega_{y\pm}$ near the level crossing, signaling the opening of a momentum bandgap.

In Fig. **2**(a), we also plot $\operatorname{Im}\Omega_{y\pm}$ for $k''_y = g/2$ calculated from Eq. (10) (symbols), which agrees well with the corresponding transfer-matrix results. Likewise, the prediction of Eq. (10) (symbols) accurately reproduces the dependence of the bandgap strength $\left|\operatorname{Im}(\Omega_{y\pm})\right|$ on $k''_y$ at fixed $\operatorname{Re}\omega_y^{(1)}$ in Fig. **2**(b). Consistent with the two-harmonic approximation underlying Eqs. (9) and (10), the Floquet mode in Eq. (8) can be approximated as

$$Y_n(t) = \left\{\hat{v}_{-1}e^{-i(\Omega_y - \omega_0)t} + \hat{v}_0 e^{-i\Omega_y t}\right\}e^{ik_y n}, \quad (11)$$

where $\hat{v}_p = \tilde{v}_p/\sqrt{|\tilde{v}_{-1}|^2 + |\tilde{v}_0|^2}$ for $p = -1, 0$, with $\tilde{v}_{-1} = P\begin{bmatrix} 0 \\ (\tilde{v}'_{-1})_2 \end{bmatrix}$ and $\tilde{v}_0 = P\begin{bmatrix} (\tilde{v}'_0)_1 \\ 0 \end{bmatrix}$.

As an illustrative example, we consider the Floquet frequency $\Omega_y = (0.5 + 0.00783i)\omega_0$, whose significance will become apparent in the subsequent waveguide analysis. Eq. (9) yields four solutions $k_{y,(\sigma)}$ with $\sigma = \pm 1, \pm 2$, namely $k_{y,(\pm 1)} = \pm 1.181 + 0.05i$ and $k_{y,(\pm 2)} = \pm 1.164 + 0.05i$, corresponding to four distinct Floquet modes $Y_{n,(\sigma)}(t)$. In Fig. **2**(c), we show the time evolution of the normalized magnitudes $\left|\hat{Y}^{s}_{1,(\sigma)}(t)\right|$, where $\hat{Y}^{s}_{1,(\sigma)}(t) = Y^{s}_{1,(\sigma)}(t)/Y^{B}_{1,(\sigma)}(0)$, on the $s = A$ (blue) and $B$ (orange) sublattices for the Floquet modes $\sigma = 1$ (left panel) and $\sigma = 2$ (right panel). The analytical approximation in Eq. (11) (symbols) accurately captures the temporal evolution obtained from the transfer-matrix calculation (solid lines), confirming the validity of the perturbative treatment in the weak-modulation regime. The modes with $\sigma = -1, -2$ (not shown) exhibit the same magnitude evolution as those with $\sigma = 1, 2$, differing only in phase.

***Skin-mode amplification driven by ghost-wave momentum bandgaps***—We now consider a Floquet waveguide obtained by truncating the AFL along the ghost-wave decay ($y$) direction. Experimentally, such a boundary in the synthetic frequency dimension can be created by strongly coupling an auxiliary resonator [53]. Since translational symmetry is preserved along $x$, the field amplitudes retains the coordinate-separable form $\psi_{n,m}(t) = Y_n(t)\exp(ik_x m - i\omega_x t)$, where $Y_n(t) = [Y_n^A(t),\quad Y_n^B(t)]^T$ is the reduced unit-cell amplitudes. The dynamics of the truncated lattice are then governed by Eq. (2), supplemented by the open-boundary conditions $Y_0(t) = Y_{N_r+1}(t) = 0$, where $N_r$ is the number of unit cells along the $y$ direction. In the isotropic stage $q = 2$ with $g(t) = g^{(2)} = 0$, the eigenvalue problem associated with Eq. (2) reduces to that of an isotropic column with open boundaries, whose eigenfrequencies and eigenmodes are $\omega^{(2)}_{y,l} = 2t_0 \cos\frac{l\pi}{2N_r+1}$ and $V^{(2)}_l = \left[\sin\frac{l\pi}{2N_r+1}, \cdots, \sin\frac{2N_r l\pi}{2N_r+1}\right]^T$, respectively, with $l = 1, \cdots, 2N_r$. In the anisotropic stage $q = 1$ with $g(t) = g^{(1)}$, the corresponding non-Hermitian SSH column is related to the isotropic counterpart through the non-unitary similarity transformation $S$. Accordingly, the eigenmodes are related by $V^{(1)}_l = SV^{(2)}_l$, where $S$ is a diagonal matrix with entries $1, e^{-g^{(1)}}, e^{-g^{(1)}}, e^{-2g^{(1)}}, \cdots, e^{-N_r g^{(1)}}$. The eigenfrequencies remain unchanged, i.e., $\omega_{y,l}^{(1)} = \omega_{y,l}^{(2)}$, whereas the eigenmodes $V_l^{(1)}$ acquire an exponential envelope and become localized near one boundary, forming real-frequency non-Hermitian skin modes [54]-[56].

In the Floquet waveguide, the reduced column field-amplitude vector $Y(t) = \left[Y_1^T(t), \cdots, Y_{N_r}^T(t)\right]^T$ can be expressed as $Y(t) = \sum_l A_l(t) V_l^{(q)} = K^{(q)} A(t)$, where $q = 1, 2$ denotes the modulation stage containing time $t$, $K^{(q)} = \left[V_1^{(q)}, \cdots, V_{2N_r}^{(q)}\right]$ is the basis matrix, and $A(t) = \left[A_1(t), \cdots, A_{2N_r}(t)\right]^T$ is the state vector. Within each modulation stage, the mode amplitudes evolve as $A_l(t) \propto e^{-i\omega_{y,l}^{(q)} t}$. As in the infinite AFL, we construct a temporal transfer matrix $M = J^{(1,2)} F^{(2)} J^{(2,1)} F^{(1)}$ to describe the evolution of the state vector $A(t)$ over one temporal unit cell. Here, the propagation matrix $F^{(q)}$ is diagonal with entries $e^{-i\omega_{y,l}^{(q)} T^{(q)}}$, and the matching matrices satisfy $\left[J^{(1,2)}\right]^{-1} = J^{(2,1)} = \left(K^{(2)}\right)^{-1} K^{(1)}$. Solving the eigenvalue problem $M|\lambda_l\rangle = \lambda_l |\lambda_l\rangle$, with $\lambda_l = e^{-i\Omega_{y,l} T}$, yields the Floquet frequencies $\Omega_{y,l}$ characterizing the reduced dynamics along the $y$ direction.

Notably, despite the purely real eigenfrequencies $\omega_{y,l}^{(q)}$ in each modulation stage, the Floquet frequencies $\Omega_{y,l}$ of the periodically driven waveguide can become complex. In general, they depend not only on the Floquet modulation itself but also on the spatial boundary conditions and the waveguide size (i.e., the number of unit cells $N_r$ along the $y$ direction). For the Floquet waveguide considered here, obtained by truncating the system of Fig. **2** along the $y$ direction with $N_r = 50$, we find a single amplifying Floquet mode with $\Omega_y = (0.5 + 0.00783i)\omega_0$. Interestingly, the amplifying mode can emerge for arbitrarily weak modulation, provided that the waveguide width $N_r$ is sufficiently large (see [50]). In contrast to previously studied Floquet lattices and synthetic PTC analogues [30]-[32], the amplification here arises entirely from periodic modulation, without requiring either material gain or gain-loss engineering.

The amplifying Floquet mode in the Floquet waveguide can be understood as a manifestation of the ghost-wave momentum bandgap in the corresponding infinite AFL. To make this connection explicit, we expand its reduced unit-cell amplitude $Y_n(t) = \Lambda_n(t)$ in the Floquet basis $Y_{n,(\sigma)}(t)$ of the infinite system [Eq. (8)] as $\Lambda_n(t) = \sum_\sigma c_{(\sigma)} Y_{n,(\sigma)}(t)$, where $c_{(\sigma)}$ are the expansion coefficients. Imposing the open-boundary conditions $\Lambda_0^B(t) = \Lambda_{N_r+1}^A(t) = 0$ and matching the coefficients of individual temporal harmonics, we obtain $\sum_\sigma c_{(\sigma)} \hat{v}_{p,(\sigma)}^B = 0$ and $\sum_\sigma c_{(\sigma)} \hat{v}_{p,(\sigma)}^A e^{ik_{y,(\sigma)}(N_r+1)} = 0$ for every integer harmonic order $p$. Since higher-order harmonics typically decay rapidly with increasing $|p|$, we can restrict the harmonic expansion to $-(N-1)/2 \leq p \leq (N-1)/2$, where $N$ is a sufficiently large odd integer. Within this approximation, the infinite AFL yields $2N$ solutions $k_{y,(\sigma)}$ with $\sigma = \pm 1, \cdots, \pm N$ for a given Floquet frequency $\Omega_y$. Defining the coefficient vector $\bar{c}$ with components $c_{(\sigma)}$, the boundary conditions can be written compactly as $\begin{bmatrix} \bar{\bar{v}}^B \\ \bar{\bar{v}}^A \end{bmatrix} \bar{c} = 0$, where $(\bar{\bar{v}}^A)_{p,\sigma} = \hat{v}_{p,(\sigma)}^A e^{ik_{y,(\sigma)}(N_r+1)}$ and $(\bar{\bar{v}}^B)_{p,\sigma} = \hat{v}_{p,(\sigma)}^B$. The corresponding Floquet frequencies $\Omega_y$ satisfy $\det \begin{bmatrix} \bar{\bar{v}}^B \\ \bar{\bar{v}}^A \end{bmatrix} = 0$, revealing that the Floquet spectrum—and thus the waveguide amplification—arises from the interplay between the ghost-wave momentum bandgap of the infinite AFL and the spatial boundary conditions.

For the amplifying Floquet mode $\Omega_y = (0.5 + 0.00783i)\omega_0$, the wave vectors $k_{y,(\pm 1)} = \pm 1.181 + 0.05i$ and $k_{y,(\pm 2)} = \pm 1.164 + 0.05i$ dominate the Floquet expansion under the approximation of Eq. (9). These wave vectors belong to the $k_y'' = 0.5g$ branch selected by complex-frequency excitation in the corresponding infinite AFL [dark-red curves in Figs. **1**(b, c)]. Using the approximate Floquet basis in Eq. (11), the reduced unit-cell amplitude is written as

$$\Lambda_n(t) \approx \sum_{\sigma=\pm1,\pm2} c_{(\sigma)} \left\{ \hat{v}_{-1,(\sigma)} e^{-i(\Omega_y-\omega_0)t} + \hat{v}_{0,(\sigma)} e^{-i\Omega_y t} \right\} e^{ik_{y,(\sigma)}n}, \tag{12}$$

where the expansion coefficients $c_{(\sigma)}$, determined up to an overall normalization constant, are fixed by the open-boundary conditions. Within the present approximation, retaining only the harmonics $p = -1, 0$ and the four dominant wave vectors $k_{y,(\sigma)}$ $(\sigma = \pm1, \pm2)$, the boundary-condition equation $\begin{bmatrix} \bar{\bar{v}}^B \\ \bar{\bar{v}}^A \end{bmatrix} \bar{c} = 0$ reduces to a $4 \times 4$ homogeneous linear system for $\bar{c} = [c_{(-2)}, \quad c_{(-1)}, \quad c_{(1)}, \quad c_{(2)}]^T$, where $\bar{\bar{v}}^A$ and $\bar{\bar{v}}^B$ are restricted to the rows $p = -1,0$ and columns $\sigma = \pm1, \pm2$.

In Figs. **3**(a) and (b), we consider the normalized amplifying Floquet mode $\widehat{\Lambda}_n(t) = \Lambda_n(t)/\Lambda_1^B(0)$. Figure **3**(a) plots the field intensity $\left|\widehat{\Lambda}_n^s(0)\right|^2$ at $t = 0$ as a function of the site index $ns$ $(n = 1, \cdots, N_r,\ s = A, B)$ along the $y$ direction. The exact numerical result (dark-red solid line; see also [50]) and the analytical approximation based on Eq. (12) (orange dashed line) are in good agreement, both reproducing the oscillatory exponential decay profile characteristic of ghost waves. As the number of modulation cycles increases, the field intensity at each site grows exponentially due to the Floquet amplification induced by the ghost-wave momentum bandgap, as illustrated by $\left|\widehat{\Lambda}_1^A(t)\right|^2$ in Fig. **3**(b), while preserving the relative spatial intensity profile shown in Fig. **3**(a) [see Eq. (12)].

***Reflectionless broadband pulse amplification in the Floquet waveguide***—The amplification mechanism in the Floquet waveguide is inherently broadband. Since the full Floquet frequency separates as $\Omega = \Omega_y + \omega_x$, with the real-valued $\omega_x = 2t_0 \cos k_x$, the amplification rate $\operatorname{Im}\Omega = \operatorname{Im}\Omega_y$ is independent of $k_x$. As a result, the nonresonant amplification extends across the entire Brillouin zone, $k_x \in (-\pi, \pi)$. Furthermore, since $\operatorname{Re}\Omega = \operatorname{Re}\Omega_y + \omega_x$ and, for the amplifying

Floquet mode obtained above, $\mathrm{Re}\,\Omega_y = 0.5\omega_0$, the resulting amplification covers the entire frequency interval $\mathrm{Re}\,\Omega \in (0.5\omega_0 - 2t_0, 0.5\omega_0 + 2t_0)$. This is in sharp contrast to conventional one-dimensional PTCs, where amplification is typically confined to a narrow parametric-resonance window centered around $\mathrm{Re}\,\Omega = 0.5\omega_0$ [5]. Accordingly, a pulsed excitation, which generally contains a broad distribution of momentum components, can be amplified collectively during propagation. To demonstrate this effect, we consider a Gaussian wave packet that is initially localized along the $x$ direction and whose profile along the $y$ direction is given by the skin mode $V_l^{(1)}$, with the initial column field-amplitude vector

$$\Psi_m(0) = \frac{1}{2\pi}\int_{-\pi}^{\pi} dk_x f(k_x) V_l^{(1)} e^{ik_x m}, \tag{13}$$

where $f(k_x) = (2\pi/\sigma_k^2)^{1/4} e^{-\left(\frac{k_x - k_x^0}{2\sigma_k}\right)^2} e^{-i(k_x - k_x^0)m_0}$ is the normalized spectral amplitude, $m_0$ denotes the initial wave-packet center, $k_x^0$ is the central wave vector, and $\sigma_k$ is the spectral width.

In the absence of temporal modulation, the pulse evolves freely in the static lattice corresponding to the $q = 1$ stage of the Floquet waveguide as $\Psi_m(t) = X(m,t) V_l^{(1)} e^{-i\omega_{y,l}^{(1)} t}$, where $X(m,t) = \frac{1}{2\pi}\int_{-\pi}^{\pi} dk_x f(k_x) e^{-i\omega_x t} e^{ik_x m}$ describes the free dispersive propagation along the $x$ direction. Under the narrow-band approximation, the longitudinal intensity profile is given by $|X(m,t)|^2 \approx \frac{1}{\sqrt{2\pi}\sigma_x(t)} \exp\left[-\frac{(m - m_0 - v_g t)^2}{2\sigma_x^2(t)}\right]$, where $\sigma_x(t) = \sqrt{1 + (2\sigma_k^2 \eta t)^2}/2\sigma_k$ is the wave-packet width, $v_g = -2t_0 \sin k_x^0$ is the group velocity, and $\eta = -2t_0 \cos k_x^0$ is the second-order dispersion coefficient. Under Floquet modulation, the pulse propagates along the waveguide without back-reflection while being simultaneously amplified, as illustrated by the intensity distributions at time $t = 0$, $20T$, and $40T$ in Fig. **3**(c) for the initial pulse in Eq. (13), with parameters $l = 19$, $m_0 = 700$, $k_x^0 = 0.6\pi$, and $\sigma_k = 0.06\pi$. In contrast to conventional PTCs, where amplification originates from the interference between time-refracted and time-reflected waves, the amplification mechanism demonstrated here, driven by the ghost-wave momentum bandgap, requires no such interference along the propagating $x$ direction.

In Fig. **3**(d), we consider the normalized total pulse energy $\hat{E}_{tot}(t) = E_{tot}(t)/E_{tot}(0)$ as a function of time, where $E_{tot}(t) = \sum_{n,m} \left|\psi_{n,m}(t)\right|^2$. For the pulse with the initial condition in Eq. (13), the total pulse energy exhibits exponential growth, following a short initial transient, under Floquet modulation (orange curve), whereas it remains constant in the absence of modulation (dark-red curve) because the initial transverse profile is an eigenmode of stage $q = 1$. To understand the observed asymptotic exponential growth, we expand the initial skin mode $V_l^{(1)}$ in Eq. (13) in terms of the Floquet eigenmodes of the waveguide. As modulation proceeds, the unique amplifying Floquet mode $\Lambda_n(t)$ in Eq. (12) dominates the evolution. Consequently, the field amplitude approaches $\psi_{n,m}(t) \approx X(m,t)\left[b\hat{\Lambda}_n(t)\right]$, where $b$ is the expansion coefficient associated with the normalized dominant Floquet mode $\hat{\Lambda}_n(t)$ (see [50]). The resulting approximate $E_{tot}(t)$ (circles) agrees well with the rigorous numerical result (orange curve) in the asymptotic regime. For comparison, we also plot the cases in which the Floquet modulation is terminated after several modulation cycles during stages $q = 1$ and $q = 2$. In the former case, $E_{tot}(t)$ continues to oscillate without systematic growth (green curve) owing to the nonorthogonality of the excited non-Hermitian eigenmodes, even though all the eigenfrequencies are real, whereas it remains constant in the latter case, where the lattice is Hermitian (magenta curve). Animations of the evolution of the vertically integrated pulse intensity for all the scenarios shown in Fig. **3**(d) are provided in the Supplemental Movie 1.

***Conclusions***—In this work, we have shown that ghost waves provide a new route to momentum-gap engineering in higher-dimensional Floquet lattices. By exploiting their complex-wave-vector nature together with complex-frequency excitation as an additional tuning knob, flexible momentum bandgaps can be realized without relying on material gain. Furthermore, by harnessing the higher-dimensional Floquet band structure, we realize a momentum bandgap that extends across the entire Brillouin zone along the propagation direction and a broad frequency range with arbitrarily weak modulation, allowing operation with modulation rate far below the optical carrier frequency. Leveraging this momentum bandgap, we demonstrate reflectionless broadband pulse amplification in a Floquet waveguide formed by truncating the lattice along the ghost-wave decay direction. These findings establish a higher-dimensional extension of momentum-gap physics beyond the conventional one-dimensional PTC paradigm and the space-time-duality framework, highlighting the potential of combining non-Bloch band engineering with complex-frequency excitation for wave manipulation in driven systems. Looking ahead, understanding the interplay between bulk band engineering and spatial boundary conditions may further broaden the applicability of this framework to PTCs and other Floquet wave platforms.

***Acknowledgements*** — This work was supported by the National Natural Science Foundation of China (Grant No. 12274240) and National Key R&D Program of China (2023YFA1407200).

**Figures**

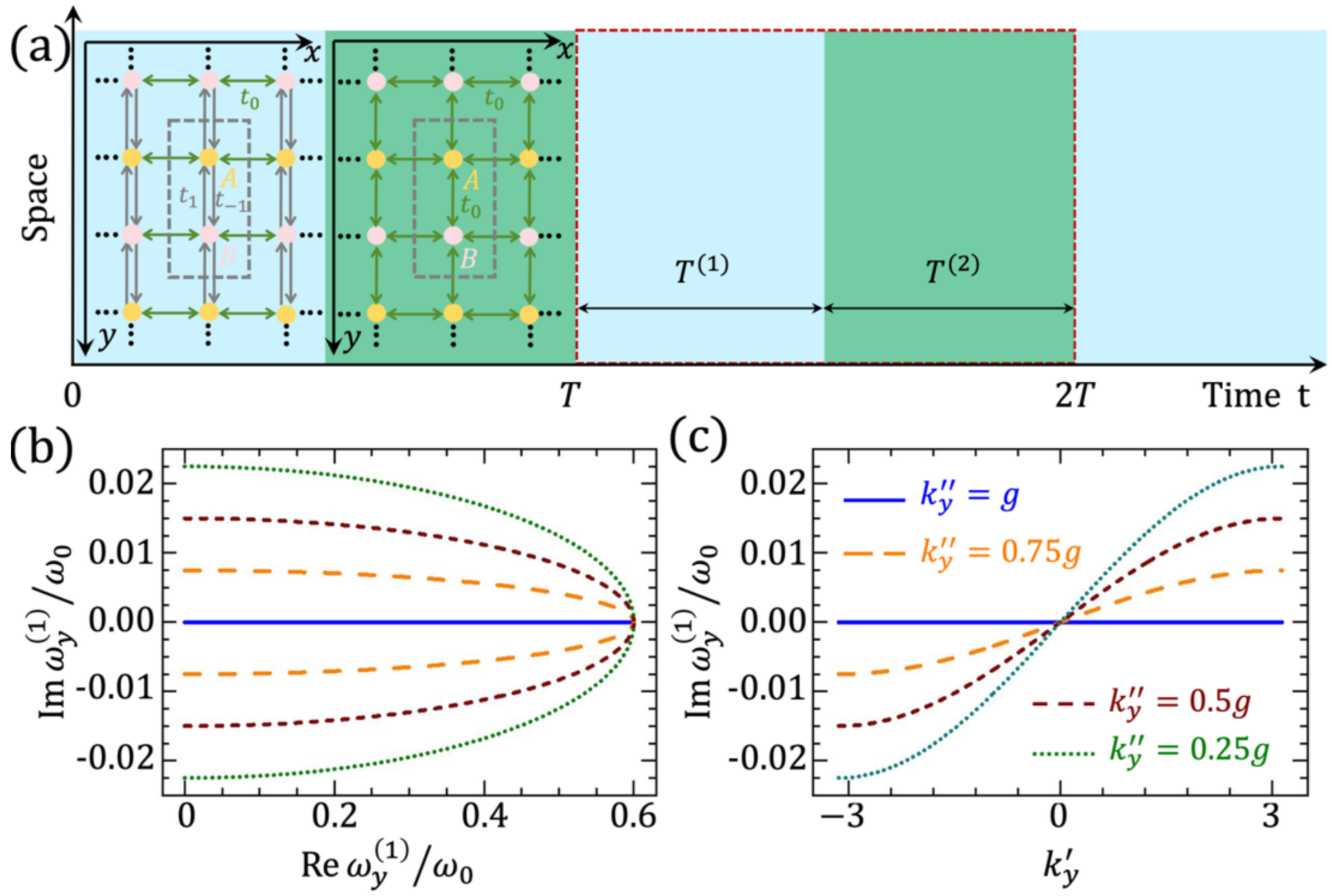

**Fig. 1.** (a) Schematic of the AFL discussed in the main text. (b) Elliptical trajectories of the complex excitation frequency $\omega_y^{(1)}$ during modulation stage $q = 1$ as the real part $k_y'$ of the complex wave vector $k_y$ is varied for fixed values of the imaginary part: $k_y'' = g$ (blue), $0.75g$ (orange), $0.5g$ (dark red), and $0.25g$ (green). (c) $\mathrm{Im}\,\omega_y^{(1)}$ as a function of $k_y'$ for the four cases shown in (b). The parameters are $g = g^{(1)} = 0.1$ and $t_0 = 0.3\omega_0$ for both (b) and (c).

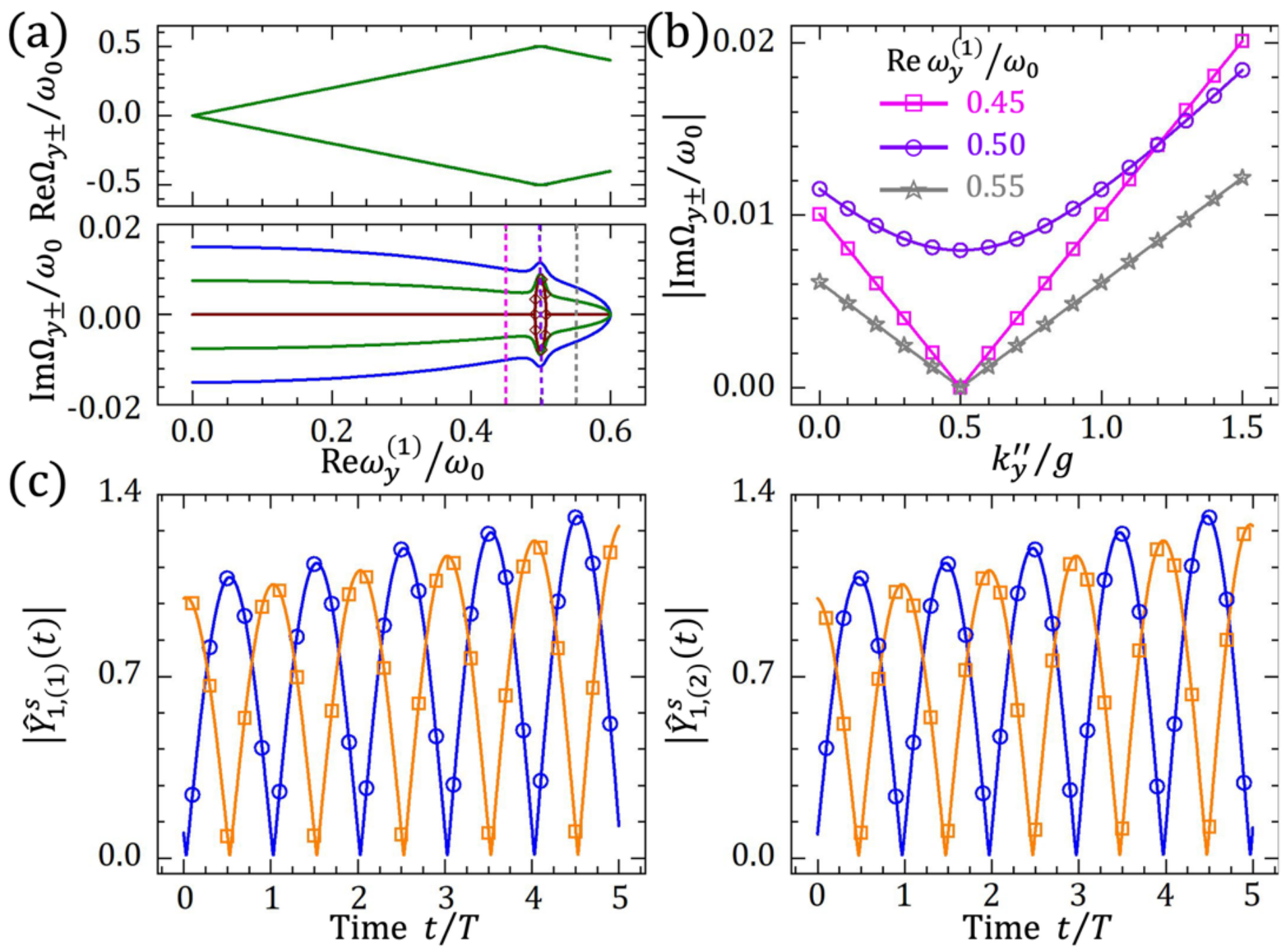

**Fig. 2.** (a) Band structure of the infinite AFL, showing the complex Floquet frequencies $\Omega_{y\pm}$ as functions of $\mathrm{Re}\,\omega_y^{(1)}$ for the four values of $k_y''$ considered in Fig. **1**(b, c), using the same color scheme. The real parts, $\mathrm{Re}\,\Omega_{y\pm}$ (upper panel), are nearly identical for different values of $k_y''$, while the imaginary parts, $\mathrm{Im}\,\Omega_{y\pm}$ (lower panel), coincide for $k_y'' = 0.75g$ and $k_y'' = 0.25g$. (b) $|\mathrm{Im}(\Omega_{y\pm})|$ versus $k_y''$ for three fixed values of $\mathrm{Re}\,\omega_y^{(1)}$, corresponding to the vertical dashed lines in (a). The symbols in both (a) and (b) are obtained from Eq. (10). (c) Normalized time-dependent magnitudes $|\hat{Y}^s_{1,(\sigma)}(t)|$ on the $s = A$ (blue) and $B$ (orange) sublattices of the first unit cell ($n = 1$) for two Floquet modes with the same $\Omega_y = (0.5 + 0.00783i)\omega_0$: $\sigma = 1$ (left panel) and $\sigma = 2$ (right panel), corresponding to $k_{y,(1)} = 1.181 + 0.05i$ and $k_{y,(2)} = 1.164 + 0.05i$, respectively. The symbols represent the analytical approximation in Eq. (11).

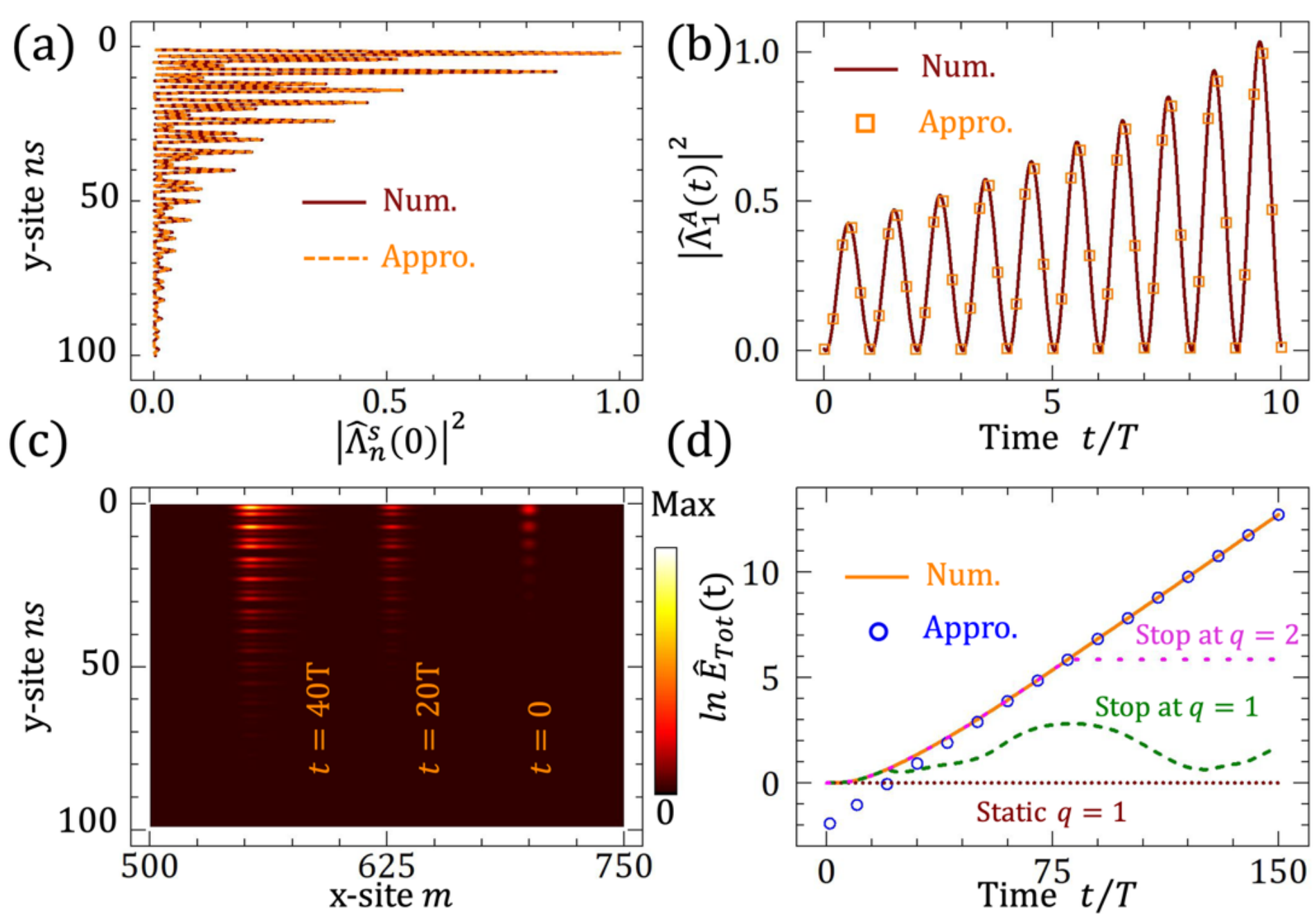

**Fig. 3.** (a) Initial profile of the unique amplifying Floquet mode in the Floquet waveguide formed by truncating the AFL in Fig. **2** along the $y$ direction with $N_r = 50$. (b) Same as (a), but for the field intensity $|\hat{\Lambda}_1^A(t)|^2$ on the $A$ sublattice of the first unit cell ($n = 1$). The symbols in (a) and (b) represent the approximation given by Eq. (12). (c) Energy distribution $|\psi_{n,m}^s(t)|^2$, normalized to the maximum value in the panel, at several different times for an initial Gaussian wave packet with parameters $l = 19$, $m_0 = 700$, $k_x^0 = 0.6\pi$, and $\sigma_k = 0.06\pi$. (d) Evolution of the normalized total pulse energy $\hat{E}_{tot}(t)$ associated with (c), obtained from the numerical result (orange line) and the asymptotic approximation (blue circles). The dark red curve represents the static lattice with $q = 1$, while the green and magenta curves correspond to the cases where the Floquet modulation is terminated after several cycles, followed by evolution with the Hamiltonian frozen at $q = 1$ and $q = 2$, respectively.